\documentclass[11pt]{article}
\usepackage{booktabs}
\usepackage{amsmath}
\usepackage{multirow}
\usepackage[final]{acl}

\usepackage{times}
\usepackage{latexsym}

\usepackage[T1]{fontenc}
\usepackage[utf8]{inputenc}

\usepackage{microtype}

\usepackage{inconsolata}

\usepackage{graphicx}

\title{IntentTune: Using user demand and personalization to resolve ``unknown'' query intents for e-commerce search}

\author{Rachith Aiyappa, Ishita Khan, Chester Palen-Michel, Jayanth Yetukuri, \\  \bf{Samarth Agrawal, Mehran Elyasi, Shuang Zhou} \\
        eBay Inc., USA}

\begin{document}
\maketitle
\begin{abstract}
Understanding user intent is fundamental to delivering relevant search results in e-commerce. 
However, substantial fraction of real-world queries are under-specified (e.g., ``watch'' or ``shirt''), lacking explicit attributes such as gender or age group. 
This ambiguity poses a significant challenge for query intent detection models in e-commerce search systems, which must accurately infer latent user intent (e.g., age, gender) to support effective downstream retrieval.
We introduce \textbf{IntentTune}, a framework for resolving ambiguous or under-specified query intents by leveraging either (1) user-specific behavioral signals including search history, browsing activity, and profile attributes or (2)  population-level demand patterns aggregated across all users. 
Through experiments on real-world e-commerce data, we first demonstrate that population-level demand patterns alone are insufficient to reliably infer intent in under-specified queries. 
We then demonstrate that user-specific behavioral signals ---particularly prior search queries --- outperform both population-level statistics and static profile information for inferring gender, age group, product category, and size intent from underspecified queries.
% Our results highlight that personalization not only improves intent detection but also reduces search friction and enables a more context-aware, individualized shopping experience. 
% IntentTune provides a scalable approach to intent resolution that can be deployed across diverse e-commerce taxonomies and user segments which can reduce search friction, and enable a more context-aware and individualized shopping experience.
\end{abstract}

\section{Introduction}
\label{sec:introduction}

Retrieving relevant products for a user query is a central problem in e-commerce search. 
Modern search engines typically rely on two complementary families of retrieval approaches: 
(i)~keyword-based lexical retrieval methods such as inverted indexing and query-document lexical matching~\cite{robertson1994okapi}, and (ii)~embedding-based retrieval (EBR) methods that map queries and products into a shared semantic vector space~\cite{lin2024enhancing, huang2020embedding, huang2013learning, kumar2021neural}. 
% Lexical methods are highly effective when query terms explicitly overlap with product descriptions, while EBR models capture semantic similarity beyond surface-level text and help address vocabulary mismatch. 
% Together, these approaches form the backbone of large-scale retrieval systems deployed in modern e-commerce platforms.
Despite their effectiveness, both paradigms struggle with the ambiguity inherent in under-specified user queries. 
A substantial fraction of real-world queries are extremely short --- often single tokens such as ``boots,'' ``watch,'' or ``shirt'' --- and omit critical attributes such as gender, age group, style, or size. 
As a result, both lexical and semantic retrieval models frequently assign such queries to ``unknown'' or ``unspecified'' intent categories, or distribute them across coarse-grained buckets that fail to reflect the user’s true intent. 
This uncertainty propagates downstream, degrading recall, ranking quality, and overall user experience.

In e-commerce settings, users exhibit strong and persistent preferences: browsing history, past purchases, saved items, and long-term category affinities provide powerful signals about what a user likely intends when issuing queries such as ``boots'' or ``sneakers.'' 
For instance, a query like ``boots'' from a user with a history of purchasing women’s ankle boots conveys a markedly different intent than the same query issued by a user who typically shops for toddler footwear or men’s work boots. 
Ignoring such personalization signals leaves substantial intent information unexploited.

We present \textbf{IntentTune}, a framework for resolving missing or ambiguous query intents produced by existing Query Understanding (QU) systems. 
IntentTune leverages two complementary sources of information: 
(1) \textit{population-level demand patterns}, which capture aggregate trends across users, and 
(2) \textit{fine-grained user-specific signals}, derived from individual browsing and interaction histories. 
This dual conditioning enables the inference of latent intent dimensions that are not expressed in the query text alone.

We focus on fashion-related queries and show that IntentTune can reliably infer gender, age group, and size (when applicable) for a large fraction of previously unresolved queries. 
We first demonstrate that models based solely on population-level demand achieve only a modest accuracy, leaving significant room for improvement. 
We then show that incorporating user-specific behavioral signals consistently matches or outperforms demand-based models across multiple intent dimensions. 
Finally we demonstrate that leveraging historical query behavior enables more accurate refinement of sub-category predictions --- a key component of the search stack --- beyond what population-level signals alone can achieve. 

Overall, IntentTune highlights an important principle in e-commerce search: \textit{query understanding should not be static or generic, but personalized, context-aware, and grounded in user behavior}. 
Our framework provides a systematic approach for integrating personalization into intent inference, enabling more accurate and user-aligned search experiences.

\section{Related Work}
\label{sec:related}

% Understanding the intent (e.g., navigational or informational, category) of a query has long been a central pursuit in search systems~\cite{chang2020query,lee2005automatic,baeza2006intention, chistova2023representation}. In conversational search, resolving ambiguity or under-specification often relies on multi-turn interactions with users~\cite{keyvan2022approach,dhole2020resolving}. 
% In contrast, traditional non-conversational search must infer intent rapidly --- typically within milliseconds --- without any opportunity for user interaction.

\paragraph{Query Understanding in E-Commerce}
Query Understanding (QU) is foundational to traditional e-commerce search, enabling the inference of latent attributes from queries~\cite{chang2020query,lee2005automatic,baeza2006intention, chistova2023representation}. 
It encompasses a diverse set of tasks, including category classification~\cite{durai2025category,cheng-etal-2024-e}, aspect-value extraction~\cite{joshi-etal-2015-distributed,papenmeier2021dataset,farzana2023knowledge,loughnane-etal-2024-explicit}, query segmentation~\cite{palen-michel-etal-2024-queryner}, and query reformulation~\cite{yetukuri2025reformulation}. 
More recently, large language models (LLMs) have been used both for direct inference and for generating datasets of aspects~\cite{blume-etal-2023-generative} and intents~\cite{tigunova-etal-2025-fabric}. 
Despite this breadth of work, most approaches focus on extracting signals directly from the query itself or assigning labels to it. 
Relatively less attention has been paid to leveraging historical user behavior to infer attributes that are not readily identifiable from the query alone.

\paragraph{Personalization in Search and Recommendation}
A substantial body of work has explored personalization in web search and recommendation systems. 
Personalized search leverages user profiles, long-term interest patterns, location, and session-level signals to re-rank retrieved documents~\cite{teevan2005personalizing, speretta2005personalized,bennett2011personallocation}. 
In parallel, collaborative filtering and latent factor models~\cite{koren2009matrix, rendle2010factorization} have demonstrated the effectiveness of user-item interaction histories for modeling preferences. 
While these approaches underscore the importance of personalization, they primarily operate \textit{after retrieval}, influencing ranking rather than retrieval itself. 
In contrast, IntentTune incorporates personalization directly into the \textit{intent inference} stage, shaping the retrieval problem at its source rather than adjusting results post hoc.

\paragraph{Implicit Attribute and Intent Identification}
Prior work on attribute identification for underspecified queries has combined knowledge graphs~\cite{luo2021alicoco2,yu2024cosmo} with signals such as past user co-clicks at both individual and aggregate levels~\cite{luo2023implicit}.
However, \citet{luo2023implicit} do not explicitly model missing attributes using personalized \textit{individual} user search histories. 
IntentTune extends this line of work by integrating per-user preference distributions, global demand signals, and LLM-based reasoning to more accurately infer attributes such as gender, age group, size, and category.

% \paragraph{Personalized Language Models}
% Personalization in dialogue systems predates modern LLM-based approaches, with earlier work focusing on adapting dialogue agents to user-specific traits~\cite{zhang-etal-2018-personalizing}. 
% More recent efforts include benchmarking implicit personas~\citep{Jiang2025PersonaMemv2TP} and developing memory architectures such as Mem0, which combines extraction and knowledge graph techniques for LLM memory~\citep{chhikara2025mem0buildingproductionreadyai}.
% Additionally, \citet{baek2024knowledge} demonstrate how user activity can be stored in an entity-centric knowledge base and leveraged alongside LLMs for personalized query suggestion. 
% IntentTune builds on these ideas in the context of e-commerce query understanding, enabling personalized interpretation of ambiguous queries at the \textit{very first stage} of the search pipeline.

Overall, IntentTune bridges two historically distinct lines of research --- Query Understanding and personalized retrieval --- by introducing a unified framework that conditions low-level intent inference on both individual user behavior and aggregate marketplace demand.

\section{Current Pipeline and IntentTune}
\label{sec:architecture}
\begin{figure*}[t]
  \centering
  \includegraphics[width=\linewidth]{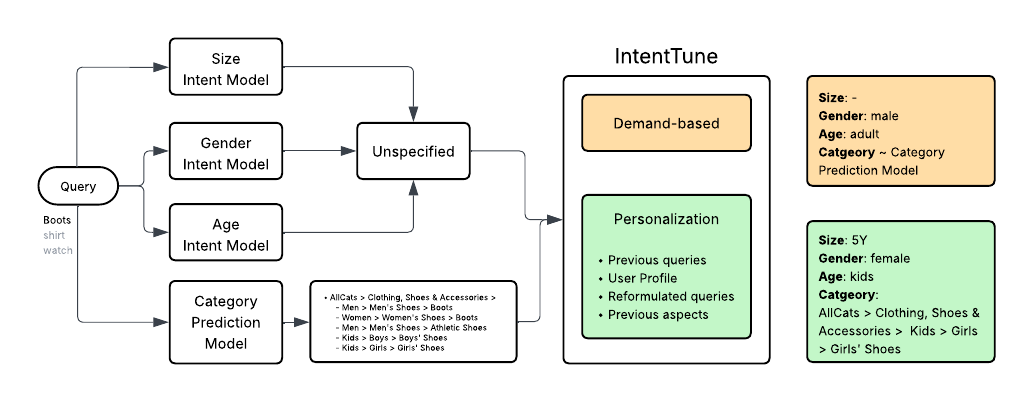}
  \caption{Overview of the IntentTune framework. 
  Queries that are labeled as ``unspecified'' by existing intent models (e.g., size, gender, and age) or insufficiently refined by category prediction models are routed to IntentTune. 
  The framework resolves such ambiguous queries by leveraging both population-level demand patterns and user-specific behavioral signals (e.g., browsing history and prior queries), enabling more accurate inference of latent intent attributes and refined category predictions.}
  \label{fig:overview}
\end{figure*}

While IntentTune is agnostic to application domains, Figure~\ref{fig:overview} illustrates how it integrates into our existing pipeline for fashion query understanding.
The overall pipeline combines semantic intent models with user-level contextual signals or population-level demand patterns to produce refined, personalized intent predictions. 
The key components are summarized
below.

\subsection{Baseline Intent Models}
Given a user query, the system first applies existing QU models to infer several core intent dimensions:
\emph{gender intent}, \emph{age-group intent}, \emph{size intent}, and \emph{category prediction}.  
Each model independently outputs either an assignment which could include an \textit{unspecified} label when the query text does not provide sufficient information. 
Many popular but coarse queries (e.g., ``nike shoes'') can fall into the ``unspecified'' bucket across one or more intent dimensions.
We note here that all our baseline intent models have a performance of greater than 0.9 (macro-F and micro-F). 

The \textbf{Size Intent Model} is a BERT-based token classification model which predicts fine-grained size entities such as \texttt{clothing\_size}, \texttt{shoe\_size}, and \texttt{size\_type} in buyer-generated queries~\cite{devlin2019bert}.
It is inherently limited to cases where size is explicitly mentioned in the query. 
For queries where size intent is implicit or omitted (e.g., ``women's sweater''), the model is unable to infer the appropriate size signal, motivating the need for additional context-aware approaches such as IntentTune.

The \textbf{Gender Intent Model} is a BERT-based classification model query into one of four interpretable classes: 
\texttt{male}, \texttt{female}, \texttt{unisex}, and \texttt{unspecified}.
Similar to the Gender Intent Model, the \textbf{Age Intent Model} is a BERT-based classification model to categorize each query into one of eight interpretable classes: 
\texttt{infant}, \texttt{toddler}, \texttt{kids}, \texttt{teen}, \texttt{young adult}, \texttt{adult}, \texttt{senior}, and \texttt{unspecified}.
Since both the gender and age intent models rely solely on query text and does not incorporate population-level or user-specific signals, it defaults to the \texttt{unspecified} class for inherently ambiguous queries such as ``sweatshirt,'' where intent is not explicitly expressed.

The \textbf{Category Prediction Model} is also a BERT based model and maps search queries to the most suitable categories in our taxonomy. 
Unlike the size, gender, and age models, this model is trained using historical search data
% incorporating interaction signals such as clicks, purchases, and post-search engagement 
to align predictions with population preferences. 
However, the predicted categories can still be inaccurate or insufficiently personalized for ambiguous or under-specified queries. 
% since the model does not explicitly account for a user’s prior search or browsing history at inference time. 
This can lead to coarse or suboptimal category assignments which affects downstream retrieval.

When one or more of the baseline intent models returns ``unspecified,'' our pipeline invokes the IntentTune framework.

\subsection{IntentTune framework}
\subsubsection{Unspecified Intent Resolution via Demand}\label{sec:demand}
Despite its limitations, the category prediction model provides useful demand-based signals that can help resolve unspecified gender and age intents by leveraging information embedded in the predicted categories. 
For instance, a query such as ``dell shirt'' is assigned to the following two categories in our taxonomy:
\begin{itemize}
    \item \textbf{AllCats > Clothing, Shoes, \& Accessories > Men > Men's Clothing}
    \begin{itemize}
        \item \texttt{Shirts > T-Shirts}
        \item \texttt{Shirts > Polos}
    \end{itemize}
\end{itemize}
We infer gender and age by selecting the category with the highest confidence score produced by the model. 
In this example, the model assigns the highest confidence to \texttt{AllCats > Clothing, Shoes, \& Accessories > Men > Men's Clothing > Shirts > T-Shirts}, which implies a \textbf{Male} gender and \textbf{Adult} age group. 
Size-related intents, however, cannot be reliably inferred from the category hierarchy, and constraining retrieval based on demand-derived size signals may negatively impact recall. 
Therefore, we do not use this module for size inference.
While this approach provides a reasonable baseline, it relies solely on the top-ranked category prediction. 
Extending this method to incorporate multiple candidate categories for intent inference is an interesting direction for future work.

\subsubsection{Unspecified Intent Resolution via Personalization}\label{sec:personalized}

This component is designed to leverage a rich set of user-level contextual signals. 
In this work we limit ourselves to two sources:
\begin{itemize}
    \item User \textbf{Profile} Attributes: Non-sensitive information provided at account creation, such as broad age group and gender. 
    These attributes do not include their size preferences and we do not use them for size-related evaluation. 
    \item \textbf{Historical queries}: The user's past search queries, from which we consider activity within a one-month window.
\end{itemize}

These signals provide strong priors over a user’s likely intent, enabling more informed disambiguation of under-specified queries. 
% Prior work has shown that leveraging user behavior and historical interactions can significantly improve personalization and retrieval relevance~\cite{search_personalization}.
For example, as illustrated in Figure~\ref{fig:overview}, a user with a history of browsing girls' footwear is more likely to intend ``girls' boots'' when issuing the ambiguous query ``boots.'' 
By conditioning on such user-specific context, the system can infer latent intent dimensions that are not explicitly expressed in the query text.

In this work, we employ an internally hosted LLM to infer gender, and age, size, and category for ambiguous queries. 
The model is prompted with (i) definitions of the intent classes, (ii) the ambiguous query, and (iii) relevant user-specific context (historical queries or user profile attributes). 
In the case of category, we also provide the LLM with the list of categories predicted by the category prediction model to choose from.
For historical queries, we include only those within a one-month window for which the baseline intent models produce high-confidence predictions ---specifically, queries with gender intent confidence greater than 0.8 or non-\textit{unspecified} age intent confidence greater than 0.9. 
These thresholds are chosen to balance the quality and quantity of contextual signals~\cite{liu2023lost}.  
It also reflects realistic deployment scenarios where only a limited number of past interactions are available.

We leave it to future work to incorporate additional sources of user context, such as previously selected aspects, saved items, and click interactions, to further improve intent inference.

\subsection{IntentTune Outputs}
The framework outputs a set of \emph{demand-based} or \emph{personalized intent assignments}, including refined predictions for size, gender, age group, and final category. 
For example, for a user with a history of browsing girls' footwear, the personalization module in Figure~\ref{fig:overview} for the query ``boots,'' outputs:
\[
\text{Size: } 5Y, \quad \text{Gender: female}, \quad \text{Age: kids}.
\]

While the category output of the ``Demand-based'' module is the same as the category prediction model, Personalization further refines it to  (\texttt{AllCats > Clothing, Shoes \& Accessories > Kids > Girls > Girls' Shoes}).

This way, IntentTune has the ability to transform ambiguous queries into fully specified, user or population  aligned intent representations, enabling more accurate downstream retrieval and ranking.

\section{Dataset}

Existing public personalization datasets are for broader domains~\cite{serdyukov2014wscd,zhao-etal-2025-personalens,wu-etal-2020-mind} or target e-commerce assistants~\cite{Bernard:2023:SIGIR}, but none of these examines personalization for item retrieval using past user queries in the e-commerce domain.
The closest to our use case is \citet{amazon2017personalsearch} but this dataset generates the queries synthetically from products rather than actual user queries. 

We therefore construct and manually annotate a dataset of user queries and past user identities to evaluate the IntentTune framework. 
From a large corpus of fashion related search queries, we first identify queries with \textit{unspecified} size, age, and gender intent (using the baseline intent models), and sample 30 such ambiguous queries to form the base query set. 
To incorporate user context, we further sample 30 users with sufficient search activity to provide meaningful behavioral signals. 
Specifically, selected users must have viewed at least 20 search result pages within a session and have participated in more than one search session over a six-month period. 
We pair each ambiguous query with each selected user, resulting in a total of $30 \times 30 = 900$ query-user pairs. 
Each pair is manually annotated to determine the intended size, age, and gender attributes, enabling evaluation of intent resolution under personalized context.

The manual annotation process incorporates multiple sources of user context, including user profile attributes and historical queries. 
In addition to past queries, users may specify aspect filters such as size, gender, or category during their search sessions. 
We include these user-selected aspects as an additional source of context to aid in disambiguating otherwise unspecified queries.
The objective of the annotation process is to assign a single gender, age, and category label to each user-query pair. 
For size, we allow for multiple labels per user--query pair. 
Table~\ref{tab:Dataset} summarizes the resulting class distributions for size, gender, and age intents derived from the annotated dataset.
We note that despite allowing for multiple labels per user---query pair, most of the pairs still have just a single annotated label.
Additionally, the size, gender, and age intents of many queries still cannot be inferred despite having access to user context reflecting the challenging nature of the problem. 

\begin{table}[h!]
\centering
\small
\setlength{\tabcolsep}{6pt}
\begin{tabular}{llrr}
\toprule
\textbf{Dimension} & \textbf{Class} & \textbf{Support} & \textbf{\%} \\
\midrule
\multirow{6}{*}{\textbf{Age}} & Adult & 762 & 84.67\% \\
 & Unspecified & 107 & 11.89\% \\
 & Young Adult & 15 & 1.67\% \\
 & Kids & 15 & 1.67\% \\
 & Infant & 1 & 0.11\% \\
\cmidrule{2-4}
 & \textit{Total} & 900 & 100\% \\
% \addlinespace
\midrule

\multirow{35}{*}{\textbf{Size}} & Unspecified & 707 & 73.72\% \\
 & XXL & 48 & 4.97\% \\
 & 8 & 19 & 1.97\% \\
 & XL & 17 & 1.76\% \\
 & 12 & 17 & 1.76\% \\
 & L & 13 & 1.35\% \\
 & 8.5 & 13 & 1.35\% \\
 & 7 1/2 & 12 & 1.24\% \\
 & 7 3/8 & 12 & 1.24\% \\
 & M & 10 & 1.04\% \\
 & Small & 9 & 0.93\% \\
 & Large & 9 & 0.93\% \\
 & 3XL & 8 & 0.83\% \\
 & 52 & 6 & 0.62\% \\
 & US 9.5 & 6 & 0.62\% \\
 & 9 & 6 & 0.62\% \\
 & 42 & 5 & 0.52\% \\
 & 7 & 5 & 0.52\% \\
 & 11 & 5 & 0.52\% \\
 & 13 & 5 & 0.52\% \\
 & 36 & 4 & 0.41\% \\
 & S & 4 & 0.41\% \\
 & 16 & 4 & 0.41\% \\
 & 9.5 & 4 & 0.41\% \\
 & 10 & 4 & 0.41\% \\
 & 15.5 & 3 & 0.31\% \\
 & 38 & 3 & 0.31\% \\
 & Boys Small & 2 & 0.21\% \\
 & Small Petite & 1 & 0.10\% \\
 & 15.15 & 1 & 0.10\% \\
 & 2C & 1 & 0.10\% \\
 & US 4--10 & 1 & 0.10\% \\
 & US 4--16 & 1 & 0.10\% \\
\cmidrule{2-4}
 & \textit{Total} & 965 & 100\% \\
% \addlinespace
\midrule

\multirow{5}{*}{\textbf{Gender}} & Unspecified & 362 & 40.22\% \\
 & Male & 354 & 39.33\% \\
 & Female & 183 & 20.33\% \\
 & Unisex & 1 & 0.11\% \\
\cmidrule{2-4}
 & \textit{Total} & 900 & 100\% \\
\addlinespace
\bottomrule
\end{tabular}
\caption{Class distribution across the intent dimensions in the human judgment set.}\label{tab:Dataset}
\end{table}

\section{Evaluation}
To evaluate the performance of the IntentTune framework
% demand-based, profile-based personalization, and historical-query-based personalization 
for resolving size, age, and gender intents, we report accuracy, weighted precision, weighted recall, and weighted F1-score. 
The weighted variants of these metrics account for class imbalance by averaging per-class performance proportional to the number of instances in each class, providing a more reliable estimate of overall performance in skewed distributions (Table~\ref{tab:Dataset}). 
In the case of size we are only evaluating the historical-query based personalization (\S\ref{sec:personalized} and \S\ref{sec:demand}) and reward the module if it gets atleast one of the possible size intents of a given user-query pair right.

% \paragraph{Size Evaluation Protocol}
% For the \textit{size} attribute, both predictions and ground-truth labels are treated as sets of comma-separated values, allowing for multiple valid sizes per query. Prior to evaluation, labels are normalized by mapping \textit{``Can't Infer''} to \textit{``Unspecified''}.
% A prediction is considered \textit{correct} if there is any overlap between the predicted and ground-truth label sets. Formally, for a predicted set $P$ and ground-truth set $T$, correctness is defined as:
% \[
% P \cap T \neq \emptyset
% \]

% This defines a lenient matching criterion, where partial agreement (i.e., at least one shared label) is sufficient for correctness. Based on this criterion, evaluation is reduced to a binary correctness task (correct vs.\ incorrect), and standard metrics such as accuracy, precision, recall, and F1-score are computed over these binary outcomes.

For category prediction, we do not directly evaluate the demand-based component of IntentTune, as it relies on the same underlying category prediction model. 
Our focus is not on improving category prediction itself, but rather on assessing the additional value introduced by personalization. 
Therefore, we evaluate the effectiveness of personalization by selecting the better-performing approach between profile-based and historical-query-based personalization (as determined by prior evaluation), and measuring the extent to which personalization improves category resolution. 
Specifically, we report the percentage of cases in which personalization successfully reduces the number of candidate categories produced by the demand-based model while preserving correctness.

\section{Results}

\begin{table}[t]
\centering
\small
\setlength{\tabcolsep}{4pt}
\begin{tabular}{llccc}
\toprule
\textbf{Dim.} & \textbf{Metric} & \textbf{Demand} & \textbf{Profile} & \textbf{Hist.\ Queries} \\
\midrule
\multirow{4}{*}{\textbf{Age}} & Acc & 0.674 & 0.650 & \textbf{0.833} \\
 & P$_w$ & 0.728 & 0.746 & \textbf{0.802} \\
 & R$_w$ & 0.674 & 0.650 & \textbf{0.833} \\
 & F1$_w$ & 0.698 & 0.687 & \textbf{0.816} \\
\addlinespace
\multirow{4}{*}{\textbf{Size}} & Acc & - & - & \textbf{0.743} \\
 & P$_w$ & - & - & \textbf{1.000} \\
 & R$_w$ & - & - & \textbf{0.743} \\
 & F1$_w$ & - & - & \textbf{0.853} \\
\addlinespace
\multirow{4}{*}{\textbf{Gender}} & Acc & 0.387 & 0.387 & \textbf{0.741} \\
 & P$_w$ & 0.356 & 0.460 & \textbf{0.765} \\
 & R$_w$ & 0.387 & 0.387 & \textbf{0.741} \\
 & F1$_w$ & 0.330 & 0.382 & \textbf{0.726} \\
\addlinespace
\bottomrule
\end{tabular}
\caption{IntentTune performance across intent dimensions. 
Best values per row are in bold. Metrics are weighted based on support sizes of classes belonging to each intent. Acc. is accuracy, $P_w, R_w, F_w$ are weighted precision, recall, and F respectively.}\label{tab:metrics}
\end{table}

\begin{figure}[t]
    \centering
    \includegraphics[width=1\linewidth]{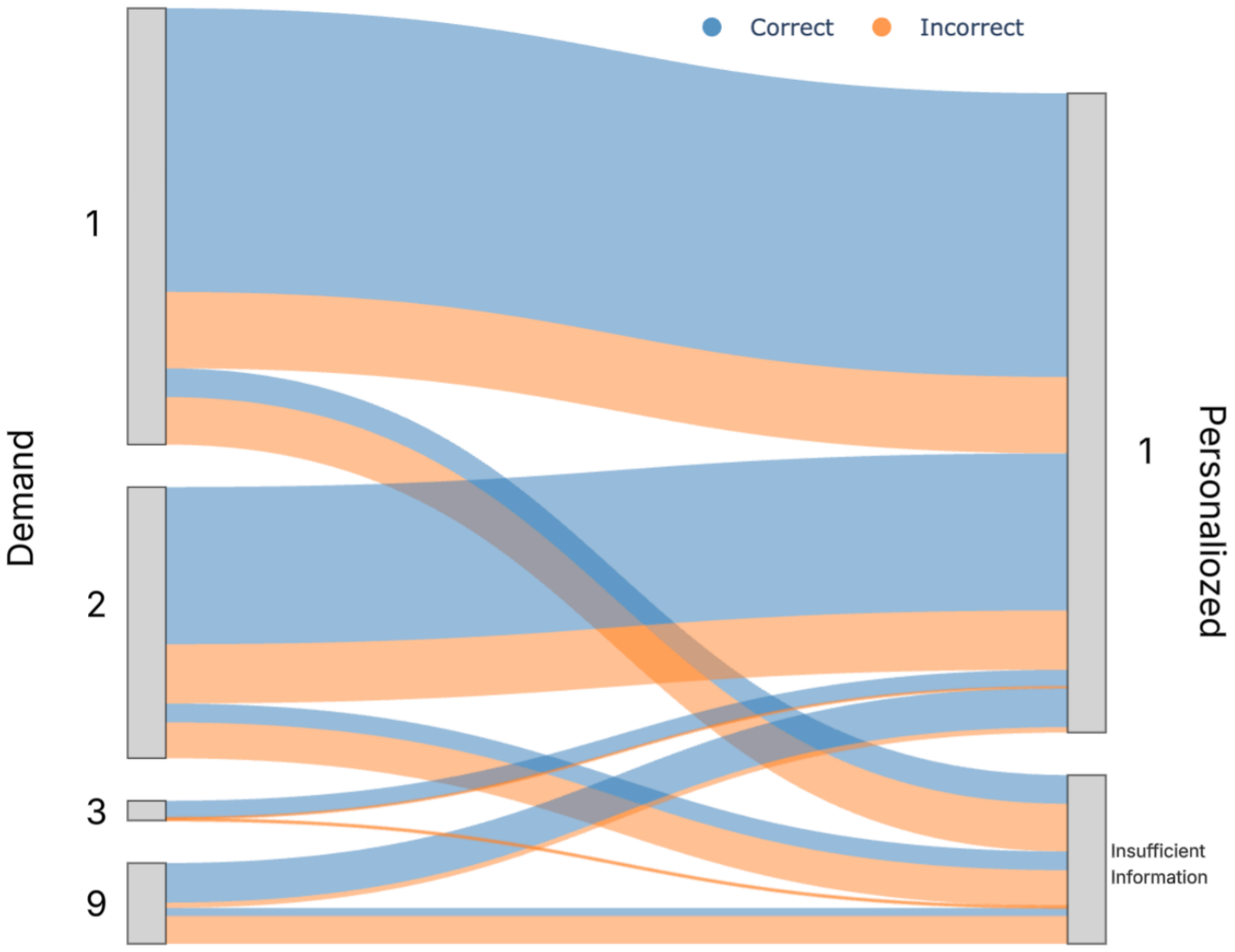}
    \caption{Historical-query-based personalization (right; ``Personalized'') refines category predictions produced by the category prediction model (left; ``Demand'') for ambiguous queries. 
    Numbers adjacent to the gray blocks indicate the number of candidate categories predicted by each module. 
    Flows represent how personalization reduces or maintains the number of candidate categories, with ``Insufficient Information'' denoting cases where personalization cannot further refine predictions due to limited user context or incorrect candidate categories. 
    Blue (orange) edges indicate cases where the refined prediction is correct (incorrect) according to manual annotation.
    }
    \label{fig:results_cat}
\end{figure}

We first observe that the demand-based module predicts age and gender intent for 77.3\% and 76.56\% of ambiguous queries, respectively.
The personalization based on historical queries (user profile attributes) predicts age and gender intent for 90.22\% (73.44\%) and 75.44\% (86.44\%) of ambiguous queries, respectively.

In Table~\ref{tab:metrics}, we compare demand-based, profile-based personalization, and historical-query-based personalization across size, age, and gender intent resolution. 
We observe that personalization based on historical queries consistently outperforms both demand-based and profile-based approaches across all evaluated dimensions.
For age, historical-query-based personalization yields substantial improvements over the demand baseline, achieving a gain of 17\% in weighted F1 score. 
% Similar gains are observed across weighted precision, recall, and F1-score, indicating that historical user behavior provides strong signals for age intent inference. 
For gender, which is a more balanced dataset, the improvements are even more pronounced. 
Historical-query-based personalization nearly doubles the accuracy relative  to both demand-based and profile-based approaches, with a gain of more than >90\% in weighted F scores. 
In contrast, profile-based personalization performs comparably to the demand-based model, suggesting that coarse-grained user attributes alone may not be sufficient for fine-grained intent disambiguation.
In other words static user attributes are less informative than dynamic behavioral context.
For size, evaluation is restricted to the historical-query-based personalization module due to the multi-valued nature of size intent and the lack of reliable size signals in demand-based and profile-based approaches. 
The results indicate that historical queries provide meaningful context for resolving size ambiguity when explicit size mentions are absent.

Fig.~\ref{fig:results_cat} shows personalization based on historical queries effectively refines category predictions produced by the category prediction model for ambiguous queries. 
Specifically, 68.5\% of the candidate categories generated by the demand-based module are correctly reduced to a single category through personalization. 
Among these cases, 10\% are flagged for further review, indicating instances where the candidate set produced by the category prediction model may require expansion or correction.

\section{Discussion}

The IntentTune framework to resolve ambiguous queries
by relying on demand-based or personalized modules
Our results demonstrate that the choice of personalization signal plays a critical role in intent resolution. 
User behavioral signals derived from historical queries are significantly more effective than both population-level demand patterns and static profile attributes. 
Our work is not focused on achieving state-of-the-art performance; rather, it serves as a proof of concept demonstrating the gains achievable by incorporating user context into query understanding. 
We leave to future work the evaluation of these improvements using online e-commerce metrics such as user satisfaction, click-through rate, and engagement. 

In sum, our work contributes to the more effective search systems by leveraging both demand-based and user-specific personalization signals.

\section{Limitations}
A central challenge in the IntentTune setting is resolving conflicting intent signals arising from multiple sources of user context, including population-level demand and user-specific information.
For example, user profile attributes may suggest an adult-oriented intent, while historical queries indicate a preference for children's items, and previously selected aspects reflect adult preferences. 
Such inconsistencies across multiple sources of context make intent inference non-trivial, requiring mechanisms that can reconcile competing signals to arrive at a coherent interpretation of user intent.
Another limitation of IntentTune in its current design is the cold-start problem, where insufficient user-level or demand-level information is available to reliably infer intent. 
This occurs, for example, when a user has little to no historical activity, when demand signals are sparse, or when queries are intentionally exploratory and not tied to prior intent. 
In such cases, the lack of contextual signals makes it difficult to provide accurate personalized predictions.

\bibliography{main}

\end{document}